\documentclass[a4paper,11pt]{article}
\usepackage{pos}
\usepackage{amsmath}

\title{Beauty hadron production in high energy proton-proton and 
heavy-ion collisions in the EPOS4HQ framework}

\author[a,b]{Jiaxing Zhao}
\author[c]{J\"org Aichelin \footnote{invited speaker}}
\author[c]{Pol Bernard Gossiaux}
\author[c]{Klaus Werner}

\affiliation[a]{Helmholtz Research Academy Hessen for FAIR (HFHF), GSI Helmholtz	Center for Heavy Ion Physics. Campus Frankfurt, 60438 Frankfurt, Germany}
\affiliation[b]{Institute for Theoretical Physics, Johann Wolfgang Goethe University, Max-von-Laue-Str. 1, 60438 Frankfurt am Main, Germany }
\affiliation[c]{SUBATECH, Nantes University, IMT Atlantique, IN2P3/CNRS
4 rue Alfred Kastler, 44307 Nantes cedex 3, France}

\abstract{Charmed hadron observables from the RHIC to LHC energies have been very successfully described with the recently advanced EPOS4HQ event generator. Here we extend this investigation to the production of beauty hadrons in proton-proton (pp) and heavy ion (HI) collisions .}

\FullConference{10th Int. Conference on ”Quarks and Nuclear Physics” (QNP-2024)\\
University of Barcelona, Spain, 8-12 July, 2024}

\begin{document}
\maketitle

\section{Introduction}

The recently advanced EPOS4HQ event generator for heavy quarks and heavy hadrons has very successfully described many charmed hadron observables in pp as well as in heavy-ion (HI) collisions \cite{Zhao:2023ucp,Zhao:2024ecc} and allowed to identify how QCD processes show up in the experimental observables. In this proceeding, we continue our studies with bottom hadrons. 

Due to their large mass, bottom quarks add information to  that obtained by charm quarks, especially because the different pQCD mechanisms contribute differently to charm and bottom quark production. However, bottom quarks are rarely produced even in high energy pp and heavy ion collisions.  Therefore the body of data is still very limited but the present run 3 at the large hadron collider (LHC) will soon improve this situation drastically. 

Bottom hadrons decays into $D$ mesons (usually called non-prompt $D$) and charmonia, such as non-prompt $J/\psi$. So, the information of bottom quark production, energy loss, and hadronization are encoded in the properties of non-prompt $D$ and non-prompt $J/\psi$.

\section{Beauty hadrons in pp}
\begin{figure}[h!]
\centering
\includegraphics[width=0.45\textwidth]{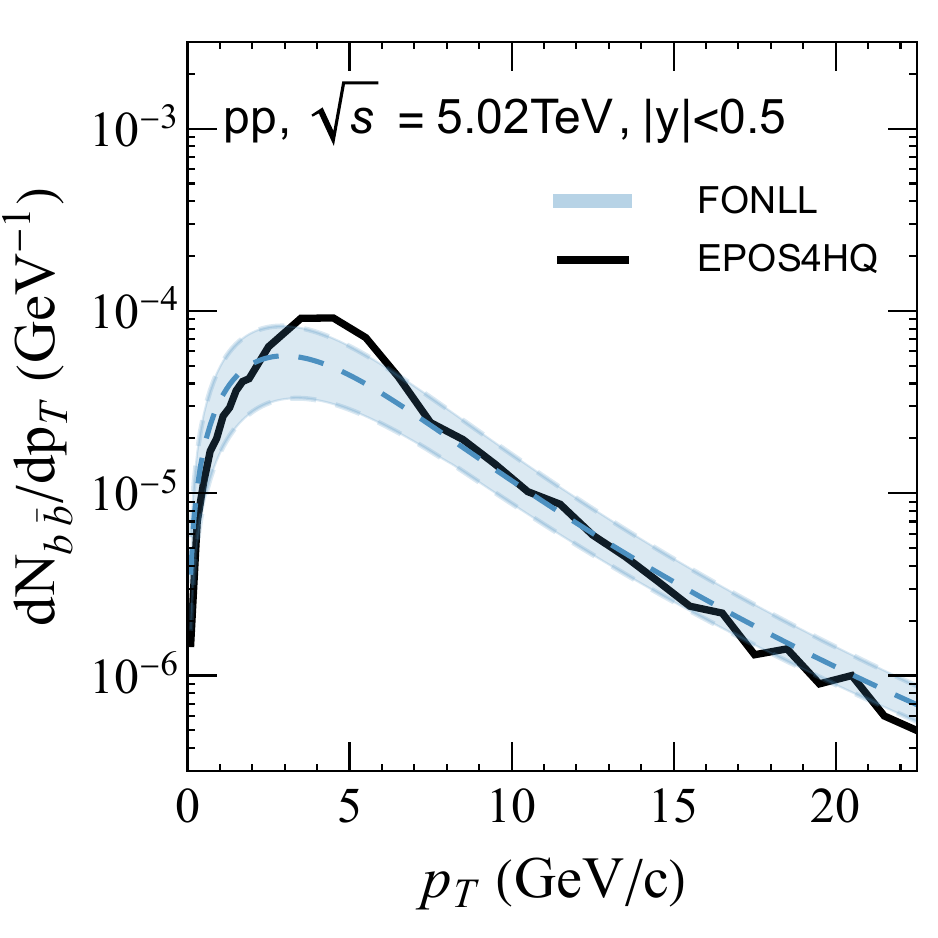}\includegraphics[width=0.45\textwidth]{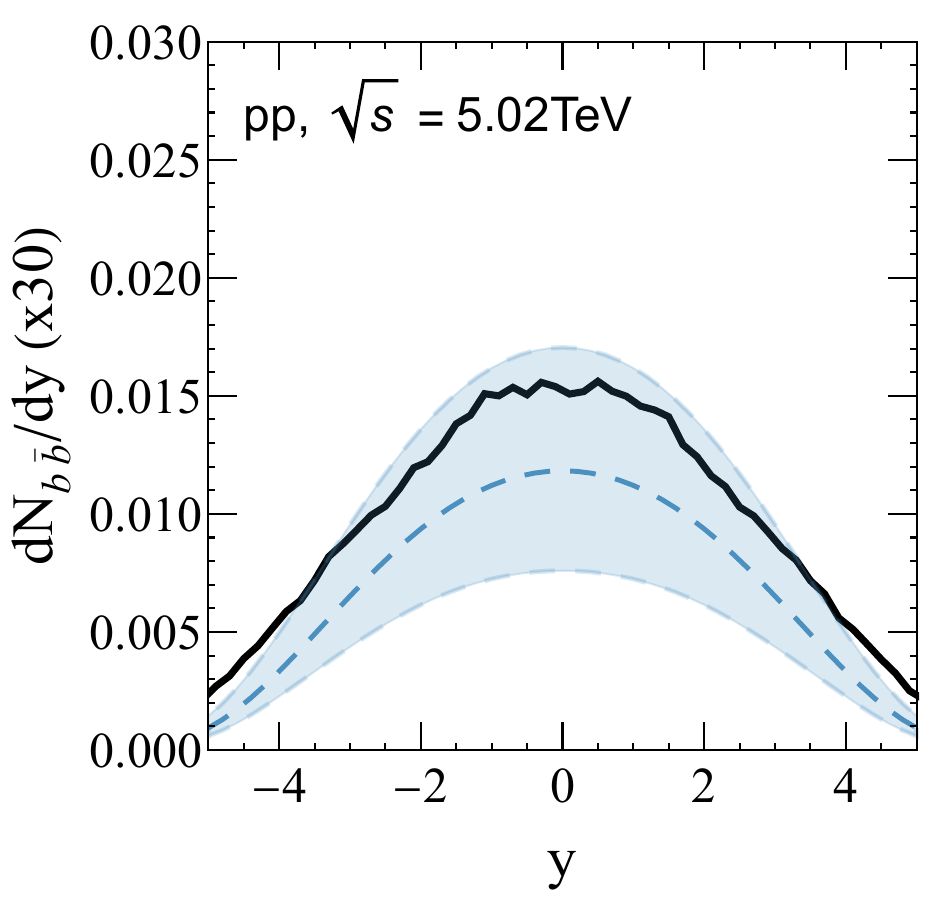}
\caption{
The transverse momentum and rapidity distribution of bottom quarks in pp collisions at $\sqrt{s}=5.02$~TeV. The black solid lines are from EPOS4HQ, while the blue bands are from FONLL calculations~\cite{Cacciari:1998it}.}   
\label{fig.bquark.pp}     
\end{figure}

In EPOS4, $Q\bar Q$ pairs can be created in three leading order or next to leading order processes. There are gluon splitting, flavor creation, and flavor excitation. In addition to these basic processes, higher order terms contribute as well, due to the realization of full space-time cascades, see detail in Ref.~\cite{Werner:2023fne,Zhao:2024ecc}
The transverse momentum, $p_T$, spectrum at midrapidity and the rapidity distribution of bottom quarks at production in pp collisions in EPOS4HQ  are shown in Fig.~\ref{fig.bquark.pp}  and compared with the prediction of FONLL \cite{Cacciari:1998it}, the most advanced pQCD calculations. We see that the EPOS4HQ predictions agree quite nicely within the error bars with the FONLL results. 
In EPOS4HQ in space regions with an energy density of light partons $\epsilon > 0.57 \rm GeV/fm^3$ a quark-gluon plasma (QGP) is created, in pp as well as in HI collisions. If the heavy quarks are in a region where a QGP exists, they interact with the QGP degrees of freedom, either elastically or inelastically. These interactions provoke an energy loss of the heavy quarks and have been described extensively in Ref.~\cite{Zhao:2023ucp}. 
\begin{figure}[h!]
\centering
\includegraphics[width=0.45\textwidth]{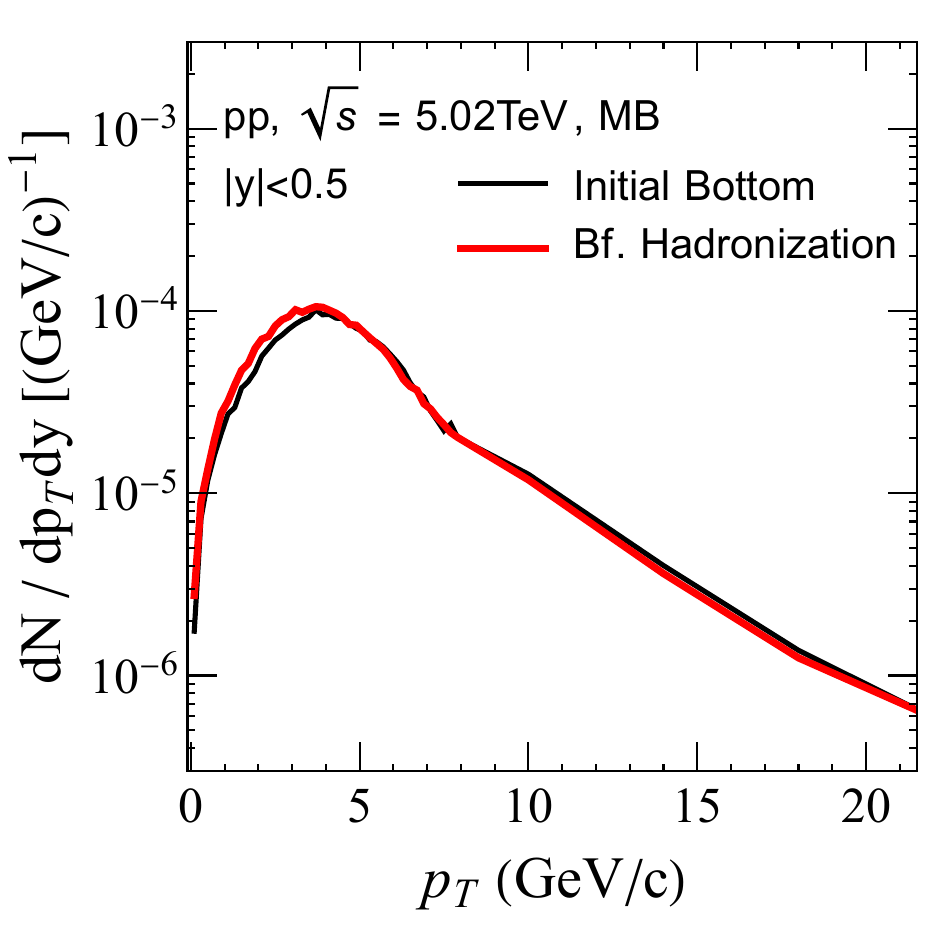}
\includegraphics[width=0.45\textwidth]{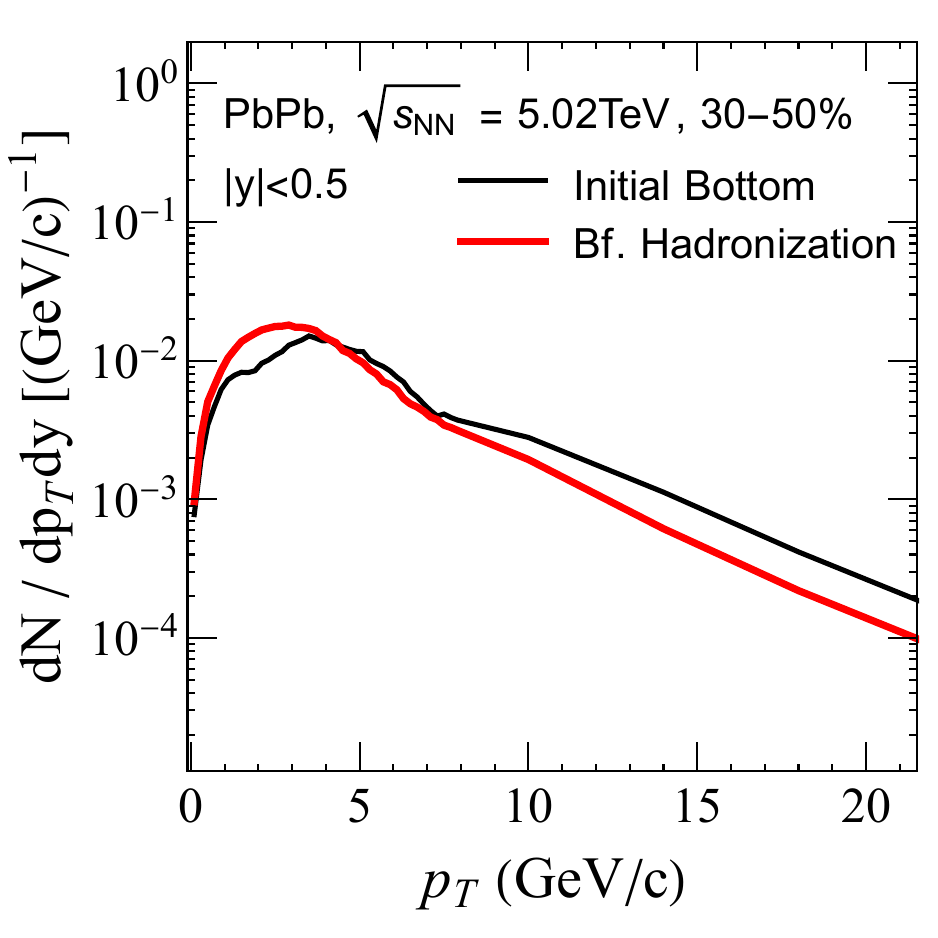}
\caption{
The $p_T$ spectra of bottom quarks at  creation and before hadronization (after energy loss in the QGP) for minimum bias (MB) pp (left) and 30-50\% centrality Pb-Pb (right) collisions at $\sqrt{s_{\rm NN}}=5.02~\rm TeV$.}   
\label{fig.ppandAA}     
\end{figure}
Due to their large mass the energy loss of bottom quarks in pp collisions is, however, tiny. This is shown in Fig.~\ref{fig.ppandAA}, left, where the $p_T$ spectra of $b$ quarks at creation and before hadronization. For HI collisions, displayed on the right side, the energy loss in the QGP is not negligible and leads to a clear modification of the $p_T$ spectra.

The hadronization of bottom quarks has been described in the above reference as well~\cite{Zhao:2024ecc}. For those quarks, which are in the QGP when it hadronizes, it is  a mixture of coalescence and fragmentation. Partons outside the QGP hadronize by pure fragmentation. The probability that a bottom hadron is formed is given in the coalescence approach by the Wigner density of a Gaussian form whose width has been adjusted to the root-mean-radius (rms) of the bottom hadron wave function, calculated with a Schrödinger equation. The details are discussed in Ref.~\cite{Zhao:2024ecc}. 

\begin{figure}[h!]
\centering
\includegraphics[width=0.9\textwidth]{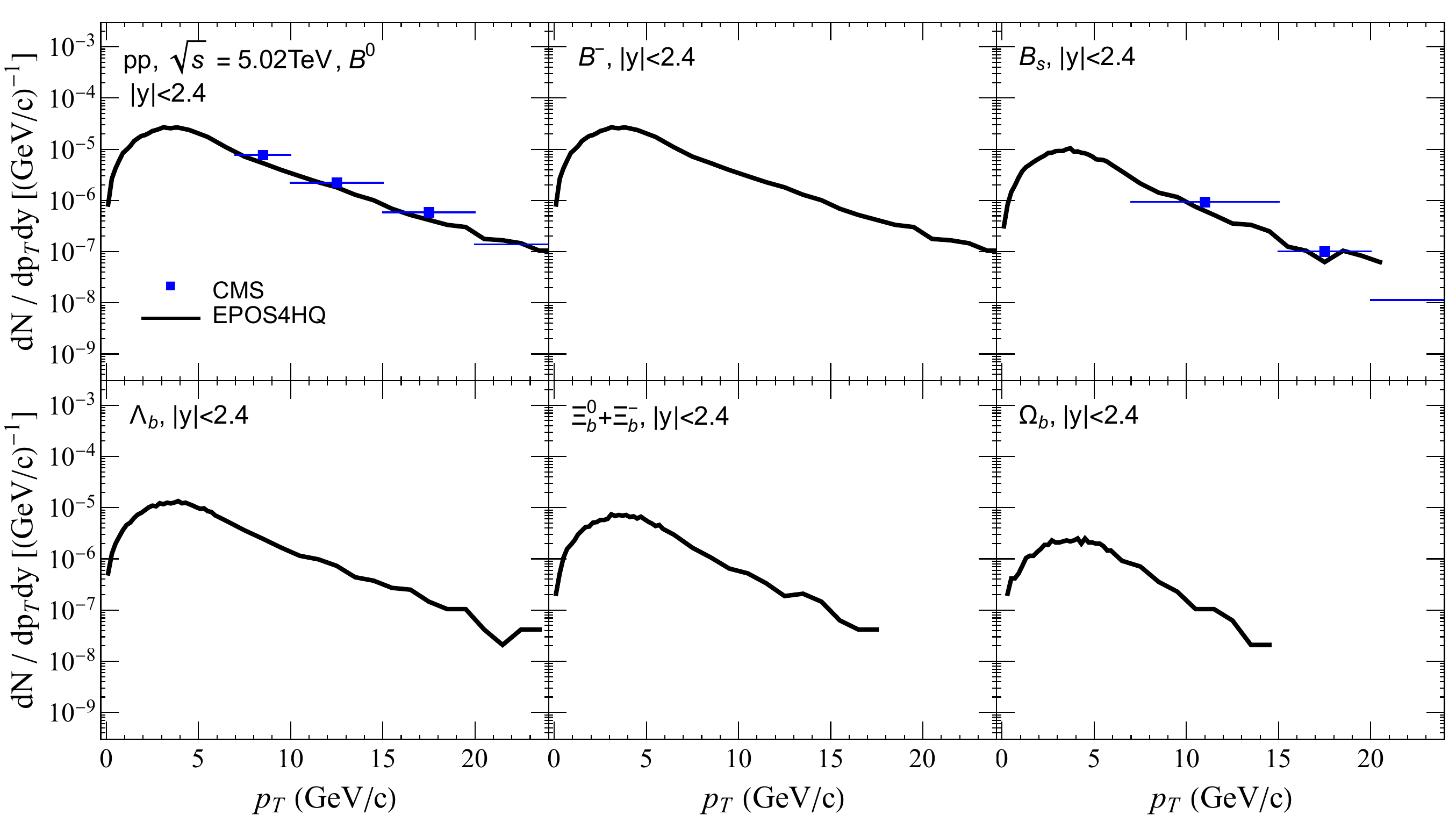}
\caption{
The bottom hadron production in minimum bias pp collisions at $\sqrt{s}=5.02$~TeV. The experimental data is from the CMS~\cite{CMS:2017uoy,CMS:2018eso}.}   
\label{fig.B.pp}     
\end{figure}

The $p_T$ spectra for $\vert y \vert < 2.4$ of the different bottom hadrons created in pp collisions at 5.02 TeV are displayed in Fig.~\ref{fig.B.pp}. In the top row we display that from $B^0, B^-$, and $ B_s$, in the bottom row that for the baryons $\Lambda_b$, $\Xi_b^0+\Xi_b^-$, and $\Omega_b$. Data exist only for $B^0$ and $B_s$ and they agree quite well with our calculations, in the slope as well as in the absolute magnitude. 
\begin{figure}[h!]
\centering
\includegraphics[width=0.45\textwidth]{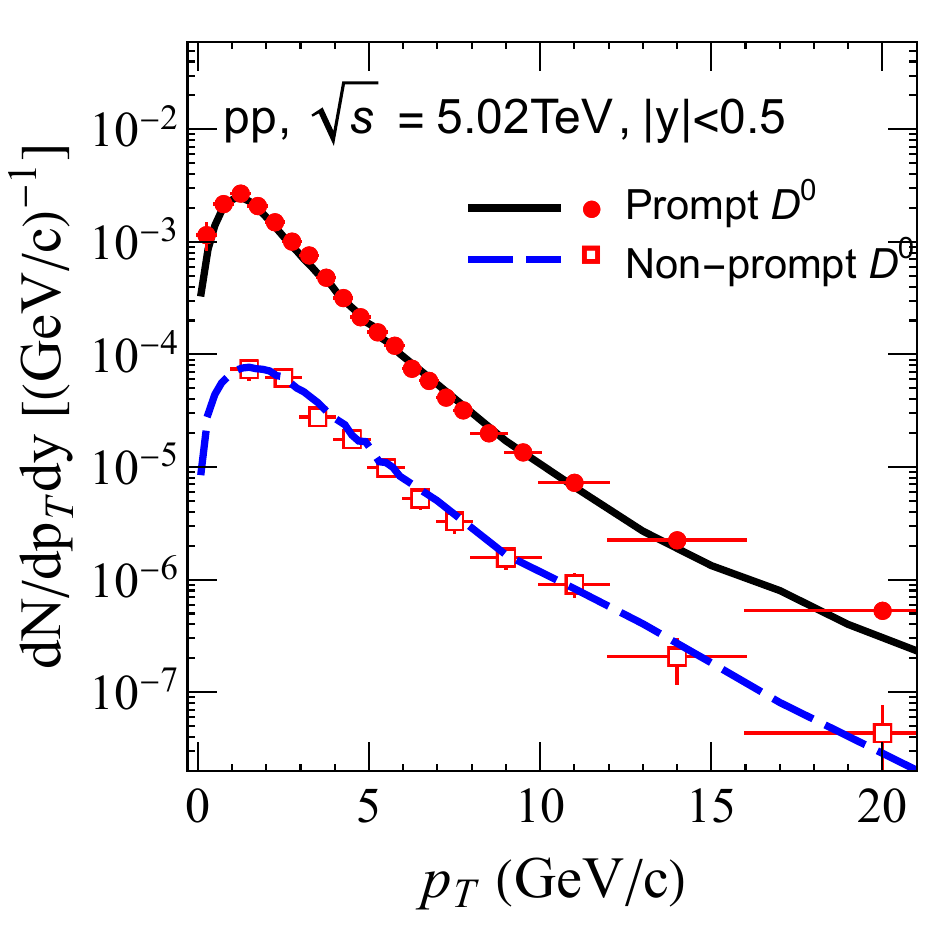}
\includegraphics[width=0.45\textwidth]{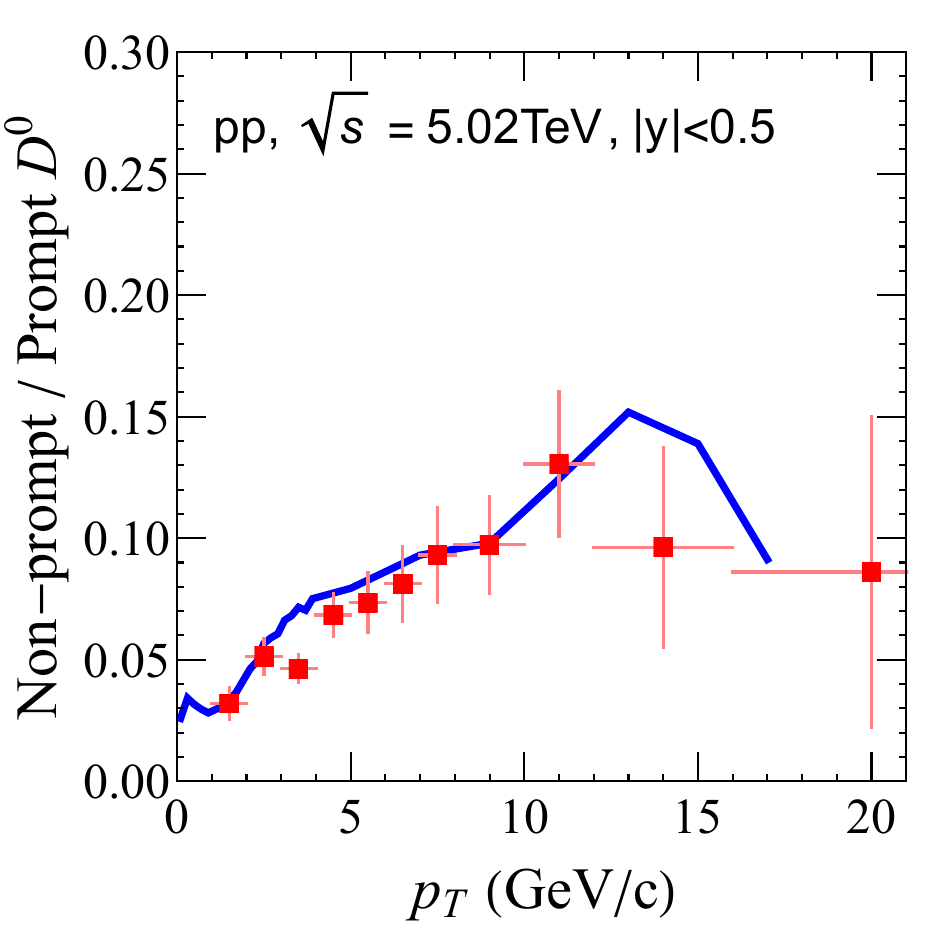}
\caption{
The prompt and non-prompt $D^0$ produced in pp collisions at $\sqrt{s}=5.02$~TeV (left) and the ratio of both (right). The experimental data is from the ALICE~\cite{ALICE:2021mgk}.}   
\label{fig.nonpD.pp}     
\end{figure}
After production bottom mesons will weakly decay into $D$-mesons. Following the PDG~\cite{PhysRevD.110.030001}, we found that the main decay channels to $D^0$ are: $B^0 \to D^0X$ with a total branching ratio $\sim51\%$, $B^-\to D^0X$ with a total branching ratio $\sim62\%$. This means that roughly half of the $B^0$ (or $B^-$) mesons decay into $D^0$. This underestimates the experimental data of non-prompt $D^0$. Considering also the other decay channels, as in this study, we obtain a larger branching ratio: 82\%, which is also used by FONLL~\cite{Cacciari:1998it}.
These $D$-mesons are called non-prompt in contrast to the directly produced $D$-meson which are formed from c quark produced at the interaction point. The ALICE collaboration has measured both contributions at midrapidity ($\vert y  \vert <$ 0.5)~\cite{ALICE:2021mgk}. The result and the comparison with the EPOS4HQ calculations is shown in Fig.~\ref{fig.nonpD.pp}  left. We see that, compared to the $B^0$ meson spectra, the spectra are shifted during the decay to the lower $p_T$. Also here we see over the whole $p_T$ range a good agreement between data and calculations.
The right hand side of Fig. \ref{fig.nonpD.pp} shows the ratio of  non-prompt and prompt $D^0$ where one can see the quality of the agreement between calculations and experiment, which is remarkable. It reflects that the  kinematics of $b$ and $c$ quark production is well under control in the EPOS4HQ approach. The increasing ratio as a function of $p_T$ is due to a harder spectrum of non-prompt $D^0$ from the $B^0$ decay  compared to prompt $D^0$, which is produced by a softer charm distribution.

\section{Beauty hadrons in HI collisions} 
As shown in Fig.~\ref{fig.ppandAA}, right, $b$-quarks, created in semi-central HI collisions, suffer a substantial energy loss in the QGP. This has influence on the $p_T$ spectra of the finally produced bottom hadrons. Bottom hadrons have been measured in 0-100\% PbPb collisions at $\sqrt{s_{\rm NN}}=5.02 ~\rm TeV$ by the CMS~\cite{CMS:2017uoy}  and compared with the EPOS4HQ results in Ref.~\cite{Zhao:2024ecc}. In the central (0-10\%) and semi-central (30-50\%) PbPb collisions at $\sqrt{s_{\rm NN}}=5.02 ~\rm TeV$, the non-prompt $D^0$ has been observed by ALICE \cite{ALICE:2023gjj}.
\begin{figure}[h!]
\centering
\includegraphics[width=0.45\textwidth]{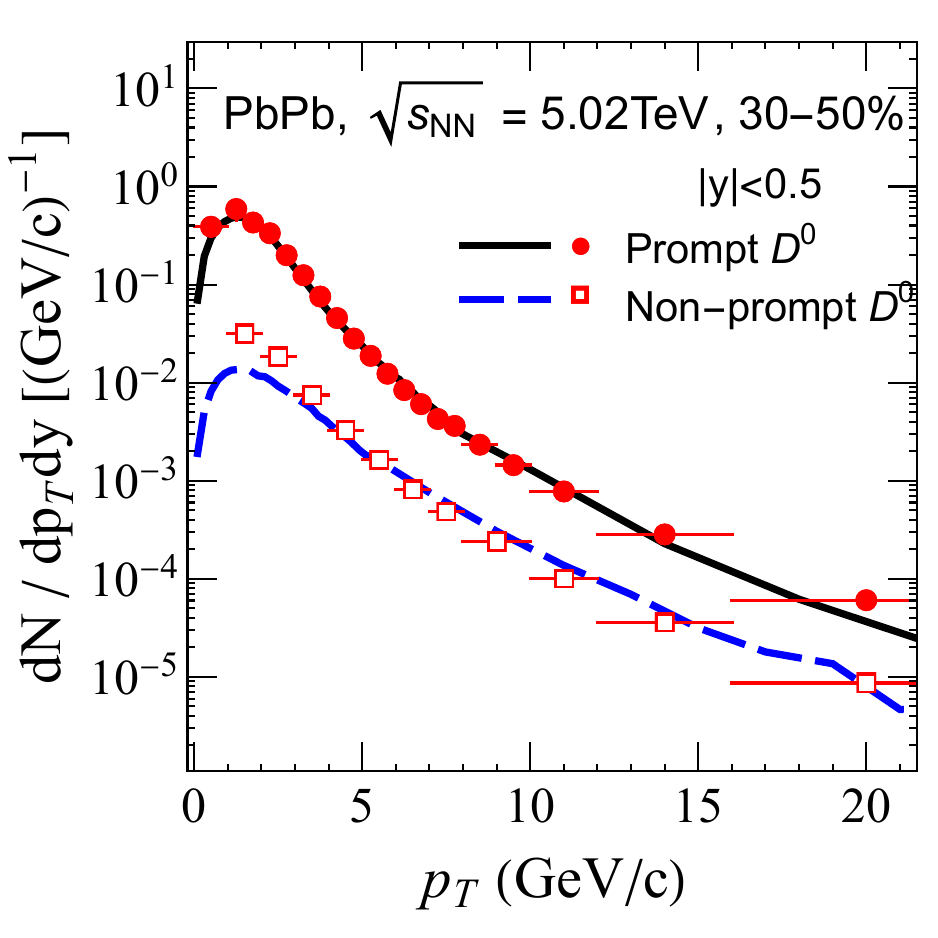}\includegraphics[width=0.45\textwidth]{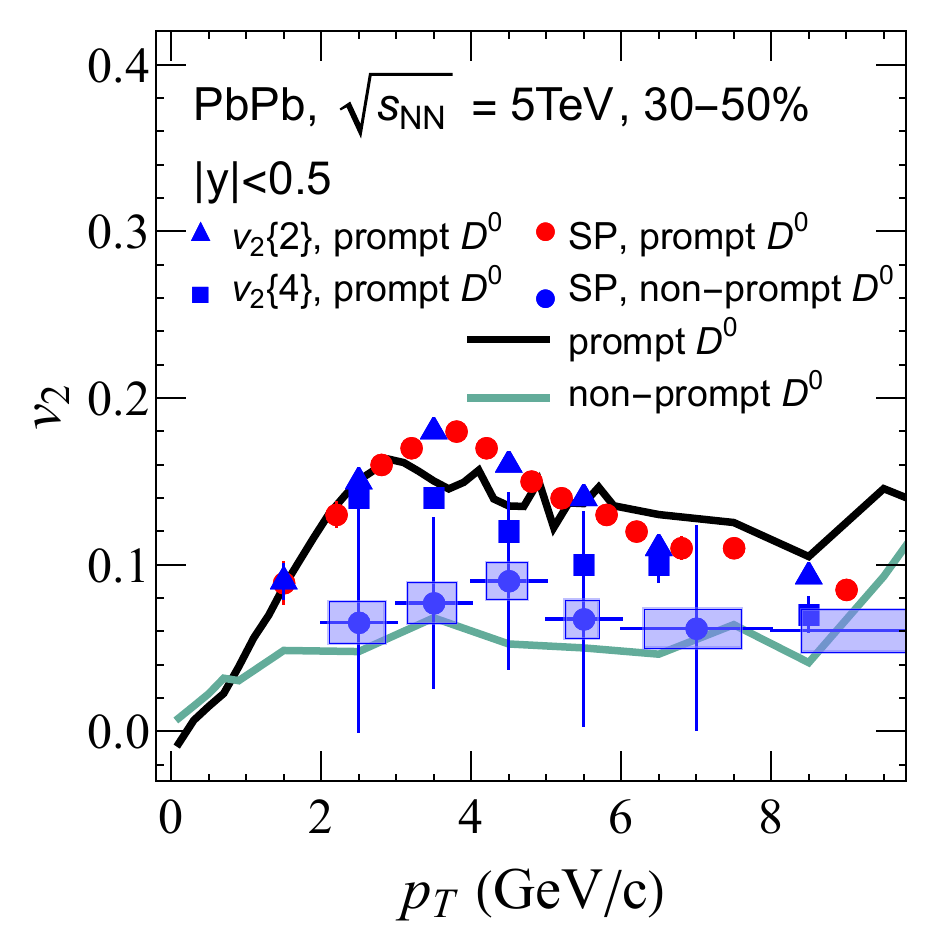}
\caption{
The prompt and non-prompt $D^0$ produced in PbPb collisions at $\sqrt{s_{\rm NN}}=5.02$~TeV. The spectra is shown in the left and the elliptic flow $v_2$ in the right panel. The experimental data are from the ALICE~\cite{ALICE:2023gjj,ALICE:2022tji,ALICE:2021rxa} and CMS~\cite{CMS:2021qqk}.} 
\label{fig.nonpD.AA}     
\end{figure}
In Fig.~\ref{fig.nonpD.AA}, left,  we compare the experimental $p_T$ spectra for prompt and non-prompt $D^0$ mesons, measured by the ALICE~\cite{ALICE:2023gjj,ALICE:2022tji} collaboration, with the EPOS4HQ results. On the right hand side we compare the EPOS4HQ elliptic flow values $v_2$  with those measured by ALICE~\cite{ALICE:2021rxa} and CMS~\cite{CMS:2021qqk}. The comparison shows that both $p_T$ spectra, that coming from $c$-quarks (prompt) and that coming from $b$-quarks (non-prompt) are well described, except a small underestimation of the non-prompt $D^0$ at the low $p_T$. The same is true for the elliptic flow $v_2$ which is half as large for $b$-decay $D^0$ compared to that of the directly produced $D^0$. Heavy quarks, produced in hard processes, show no $v_2$ when they are produced. They acquire $v_2$ by the interaction with the QGP partons. Due to different cross sections and due to the kinematics bottom quarks acquire much less $v_2$ than charm quarks. This figure shows as well that EPOS4HQ reproduces as well these flow observables.

\section{Summary}
Charmed hadron observables from RHIC to LHC energies have been very successfully described with the recently advanced EPOS4HQ event generator. In this paper, we use the same framework and investigate the production of bottom hadrons and their decay products, non-prompt $D$, in proton-proton (pp) and heavy ion (HI) collisions. We find that the experimental observables can be well described. Together with the prompt production, this allows to constrain further the heavy quark energy loss and hadronization mechanism in high-energy nuclear collisions.

\section*{Acknowledgements}
The authors acknowledge the support by the European Union's Horizon 2020 research and innovation program under grant agreement STRONG--2020 -- No 824093.

\bibliographystyle{abbrv}
\bibliography{references.bib}

\end{document}